\documentclass{iopart}
\usepackage{graphicx}

\newcommand {\pp} {pp~}

\begin{document}
\title[Femtoscopy from the ALICE experiment]{Femtoscopy of Pb-Pb and \pp collisions at the LHC with the ALICE experiment}

\author{A Kisiel$^1$ for the ALICE Collaboration}

\address{$^1$ Physics Department, CERN,  CH-1211, Gen\`eve 23, Switzerland}
\ead{Adam.Kisiel@cern.ch}

\begin{abstract}
We report on the results of femtoscopic analysis of Pb-Pb collisions
at $\sqrt{s_{NN}}=2.76$~TeV and \pp collisions at $\sqrt{s}=0.9$,
$2.76$ and $7$~TeV with identical pions and kaons. Detailed femtoscopy
studies in heavy-ion collisions at SPS and RHIC have shown that
emission region sizes (``HBT radii'') decrease with increasing pair
transverse momentum $k_{T}$, which is understood as a manifestation of
the collective behavior of matter. The trend was predicted to persist
at the LHC. The data from Pb-Pb collisions confirm the existence of a
flowing medium and provide strict constraints on the dynamical
models. Similar analysis is carried out for \pp collisions for pions
and kaons and qualitative similarities to heavy-ion data are seen,
especially in collisions producing large number of particles. The
observed trends give insight into the soft particle production
mechanism in \pp collisions. 3D radii were also found to universally
scale with event multiplicity in heavy-ion collisions. We extend the
range of multiplicities both upwards with the Pb-Pb data and downwards
with the \pp data to test the scaling in new areas. In particular the
high multiplicity \pp collisions reach particle densities comparable to
the ones measured in peripheral Cu-Cu and Au-Au collisions at
RHIC. This allows for the first time to directly compare freeze-out
sizes for systems with very different initial states.  
\end{abstract}
\submitto{\JPG}
\maketitle

\section{Introduction}
The Large Hadron Collider (LHC) has been in operation since the end of 2009
and has collided  protons at center of mass energies
$\sqrt{s}=0.9$~TeV, 2.76~TeV and 7~TeV and lead nuclei at
$\sqrt{s_{NN}}=2.76$~TeV. One of the experiments at the LHC is ALICE
which focuses on measuring heavy-ion collisions and minimum-bias \pp
collisions. The main goal of the analysis is to study the Quark-Gluon
Plasma (QGP) which is postulated to be created in Pb-Pb interactions. One of
the reasons to study \pp collisions is to provide a ``baseline'' for
the heavy-ion measurements, so direct comparisons between the two
collision types are especially interesting.

The QGP has been observed at lower energies at Relativistic Heavy-Ion
Collider
(RHIC)~\cite{Adams:2005dq,Adcox:2004mh,Back:2004je,Arsene:2004fa} and
possibly also at the Super Proton Synchrotron (SPS) at CERN. One of
the defining properties of the system created in such collisions is
its rapid evolution, which was found to be well described by the
collective models employing a hydrodynamics
framework~\cite{Kisiel:2008ws}. Specifically, strong radial expansion
(radial flow) as well as its elliptic asymmetry (elliptic flow) for 
non-central collisions have been observed and explained in the frame
of such models.

In this work we focus on a particular observable, the femtoscopic
correlation function~\cite{Kopylov:1972qw,Kopylov:1974uc}. It uses the width of the
Bose-Einstein enhancement for identical pions at low relative momentum
to measure the size of the region emitting particles. Such
measurements at lower energies in heavy-ion collisions have shown
several features~\cite{Lisa:2005dd}. Firstly, the extracted radii, in three
dimensions in the Longitudinally Co-Moving System (LCMS) scale linearly with
the cube root of the pseudorapidity density $\left <dN_{ch}/d\eta\right
>^{1/3}$. Secondly the radii decrease with increasing pair transverse
momentum, which is interpreted via the ``homogeneity lengths''
mechanism~\cite{Akkelin:1995gh} as a signature of strong collective radial
flow. Both of these effects are expected at the LHC. At the same time
femtoscopic radii have also been extracted in \pp collisions at RHIC.
The scaling between the minimum-bias \pp and heavy-ion data was
observed. We aim to test such scalings in this work in the new energy
regime. 

We note that the results for the central heavy-ion collisions have
been published by the ALICE Collaboration~\cite{Aamodt:2011mr}, while some
results for the \pp collisions have been also made
public~\cite{Aamodt:2011kd}. However the preliminary results for the \pp
collisions at $\sqrt{s}=2.76$~TeV are shown for the first time in this
work and the comparison of the \pp dataset for all $\sqrt{s}$ to the
heavy-ion collision data, including the ALICE central collision point,
is also discussed for the first time.


\section{Data analysis}

The analysis presented here was performed at the LHC with the ALICE
detector. In particular the Time Projection Chamber (TPC) was used for
tracking and particle identification, the Inner Tracking System (ITS)
was used for tracking, vertexing and trigger, and the VZERO detectors
were used for triggering and centrality determination in Pb-Pb
collisions~\cite{Aamodt:2008zz}. For the charged kaon analysis the
ALICE Time Of Flight (TOF) detector~\cite{Aamodt:2008zz} was used for
particle identification.  The \pp dataset at $\sqrt{s}=0.9$~TeV has
been collected in May 2010 and consisted of roughly 4 million events
passing the selection criteria. The dataset at 7~TeV has been
collected in June and consisted of roughly 60 million events. The 
dataset at 2.76~TeV has been collected in March 2011 and consisted of
roughly 20 million events. The Pb-Pb dataset at $\sqrt{s_{NN}}=2.76$~TeV has
been collected at the end of 2010, the subsample of approximately 16 thousand
events with centrality 0-5\% of the total inelastic cross-section was
selected for the analysis. 

\begin{figure}
\centerline{
\includegraphics[width=0.60\textwidth]{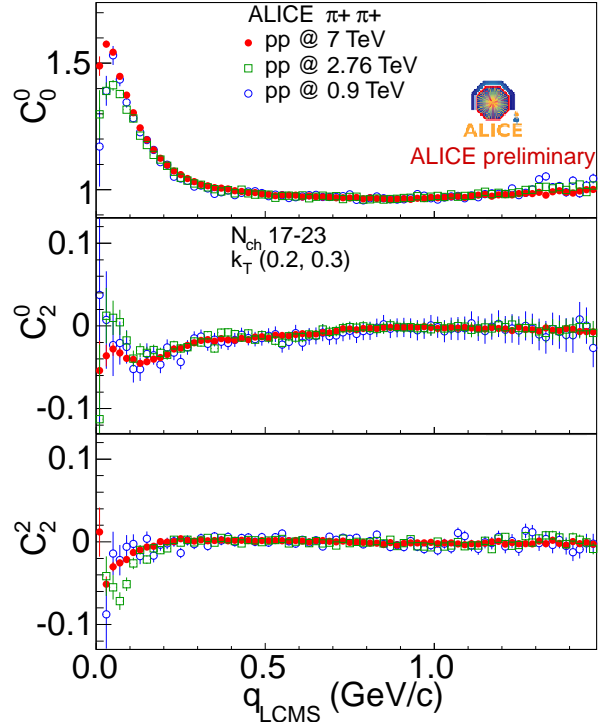}
}
\caption{Correlation functions for \pp collisions at three collision
  energies for the same multiplicity and pair momentum range. The
  three panels show the first three non-vanishing components of the
  spherical harmonics decomposition.}
\label{Figtwo}
\end{figure}

A minimum-bias trigger using the combination of the response from the
inner layer of the ITS as well as the VZERO detector was used for \pp
collisions. In this analysis, the events were also required to have an
event vertex reconstructed within 12~cm of the center of the TPC and
to have at least one pion identified. The events were categorized
according to the raw total event multiplicity in $|\eta|<1.2$, the
value for the normalized pseudorapidity density $\left <dN_{ch}/d\eta
\right >$ was also determined for each of the ranges. For Pb-Pb
collisions the events were required to have a vertex reconstructed
within 12~cm of the center of the TPC  and to be in the top 5\%
centrality bin according to the VZERO detector. 

The tracks were reconstructed with the TPC (and ITS for the \pp
analysis), and were required to pass standard criteria for a good
tracking fit. Primary particles were selected by requiring that they
point back to the primary vertex within 0.25~cm in the transverse
plane. Using the specific 
energy loss in the TPC the particles were identified as pions, with
some electrons and kaons remaining. Particles
within $|\eta|<1.2$ were accepted for \pp analysis and those within
$|\eta|<0.8$ for 
Pb-Pb collisions. The $p_{T}$ range was limited by the TPC acceptance and was
from 0.15(0.2)~GeV/c to 2.0~GeV/c for \pp(Pb-Pb) collisions.

We have also performed preliminary analysis with identical
kaons. Charged tracks with $p>0.6$~GeV/$c$ were required to be consistent
with the kaon mass hypothesis in the TOF detector to be accepted. A
small contamination from electrons remained in the sample below this
momentum. The $\mathrm{K}^0_{s}$ were identified using the ``V0'' decay topology
in the TPC from two oppositely charged pions. The resulting purity was
estimated to be 96\%.

For Pb-Pb collisions a specific two-track separation cut was used to
correct for the inefficiency of the reconstruction of two close tracks
in the detector~\cite{Aamodt:2011mr}.

\section{Correlation functions}

\begin{figure}
\centerline{
\includegraphics[width=0.60\textwidth]{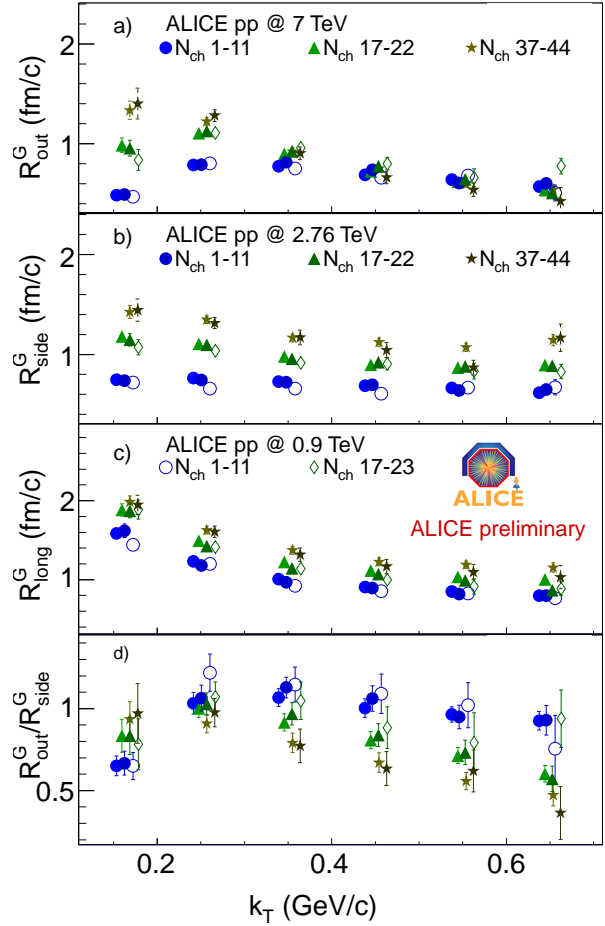}
}
\caption{Comparison of the femtoscopic radii for selected multiplicity
and pair transverse momentum ranges for different collision energies. The
points for different energies and multiplicities have been slightly
shifted in the horizontal direction for visibility.}
\label{Figthree}
\end{figure}

The identified primary pions in each event were combined into pairs to
form the ``signal'' sample $A$. The same was done for pair of pions
where each of them came from a different event to form the
uncorrelated ``background'' sample $B$. The events used for the mixing
had to have similar acceptance, so they were required to have event
vertexes within 1~cm of each other, and be in the same
multiplicity(centrality) range. The correlation functions were
constructed separately for each multiplicity/pair momentum range and
collision energy by dividing $A$ over $B$. Both $A$ and $B$ were created
as functions of the 3 components of the pair relative momentum $q$ in
LCMS, where the pair momentum along
the beam axis vanishes. Simple Cartesian representation was used for
Pb-Pb collisions while for \pp collisions the analysis was also done
in the Spherical Harmonics (SH) decomposition~\cite{Kisiel:2009iw}. An example
of the correlation function is shown in Fig.~\ref{Figtwo}. The first 
three non-vanishing SH components are shown; they are enough to
perform the fitting procedure described later.

In Fig.~\ref{Figtwo} the correlation functions are shown for all three
collision energies for the same multiplicity/pair momentum range. They
are all similar, specifically the width of the low-$q$
enhancement is comparable. Some differences in the magnitude of the
enhancement is expected due to slightly different data taking
conditions for the three samples. The correlation functions
already suggest that the emission region size is independent of
collision energy if measured for collisions with same multiplicity.

\section{Emission region sizes}

The correlation functions can be analyzed quantitatively to extract
the sizes of the emitting region. We use the assumption of the
Gaussian shape of the source which gives the following fit function:
\begin{eqnarray}
C(\vec q) &=& N [ (1-\lambda) \\
&+& \lambda K(q_{inv}) (1 +   \exp(-R_{out}^{2} q_{out}^2 - R_{side}^2 q_{side}^2 - R_{long}^2   q_{long}^2 )  ] B(\vec q) \nonumber 
\label{eq:cffit}
\end{eqnarray}
where $N$ is the normalization factor, $K$ is the Coulomb wave
function squared averaged over a Gaussian source with radius 1~fm, $q$
is the pair relative momentum, and $R$'s are the emission source
sizes. $B$ is accounting for non-femtoscopic correlations measured in
\pp collisions at the LHC which are described in detail
in~\cite{Aamodt:2011kd}. Both $q$ and $R$ have three components, the
``long'' along the beam axis, ``out'' along the pair
transverse momentum, and ``side'' perpendicular to the other two.

Figure~\ref{Figthree} shows the fitted femtoscopic radii in \pp
collisions as a function of pair momentum for three collision
energies, for selected multiplicity ranges where data from at least two
collision energies is available. At same $k_{T}$ and multiplicity the radii
do not depend on $\sqrt{s}$ which confirms the finding with raw
correlation functions. The radii universally grow with multiplicity
except the highest $k_{T}$ points for $R_{out}$. $R_{long}$ shows
a similar $k_{T}$ dependence at all multiplicities, with the overall
scale growing with multiplicity. $R_{side}$ shows a flat $k_{T}$
dependence in low multiplicity collisions, and develops a negative
slope as  it increases. For $R_{out}$ a similar but more dramatic
behavior is seen, the slope of the $k_{T}$ dependence is especially
strong at larger multiplicities. The difference in behavior between
$R_{out}$ and $R_{side}$ is most prominent in their ratio, which for
high multiplicities goes significantly below 1.0, much lower than any
value observed in heavy-ion collisions. The qualitative behavior of
radii, i.e. their decrease with $k_{T}$, is similar to the one observed
in heavy ions, however the details, especially the strong difference
in the transverse radii behavior, show that the radial flow
interpretation valid in heavy ions is not necessarily applicable to
\pp collisions. More development on the theory side is needed to give a
definite answer.

\section{Multiplicity dependence}

\begin{figure}
\centerline{
\includegraphics[width=0.48\textwidth]{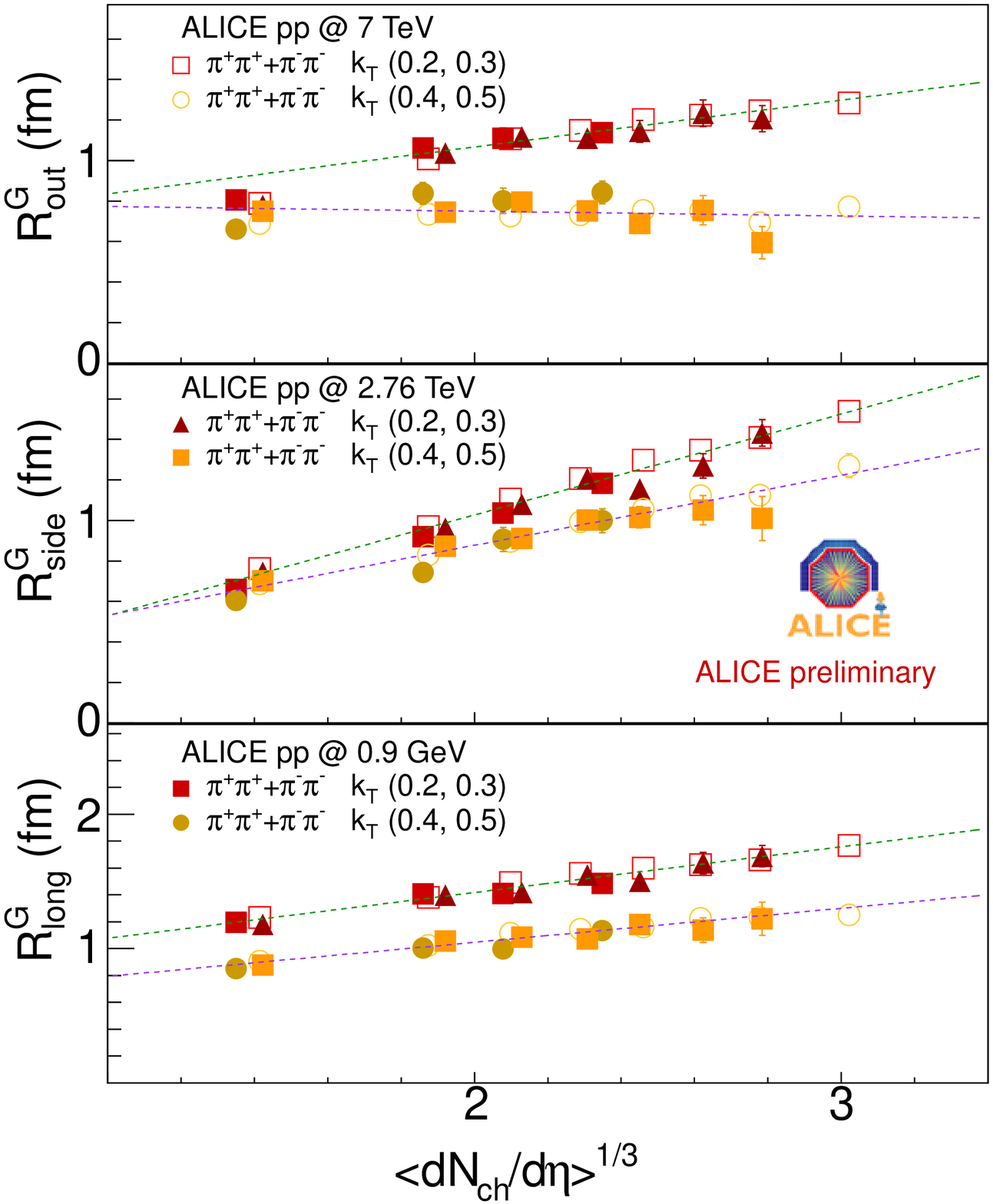}
\includegraphics[width=0.48\textwidth]{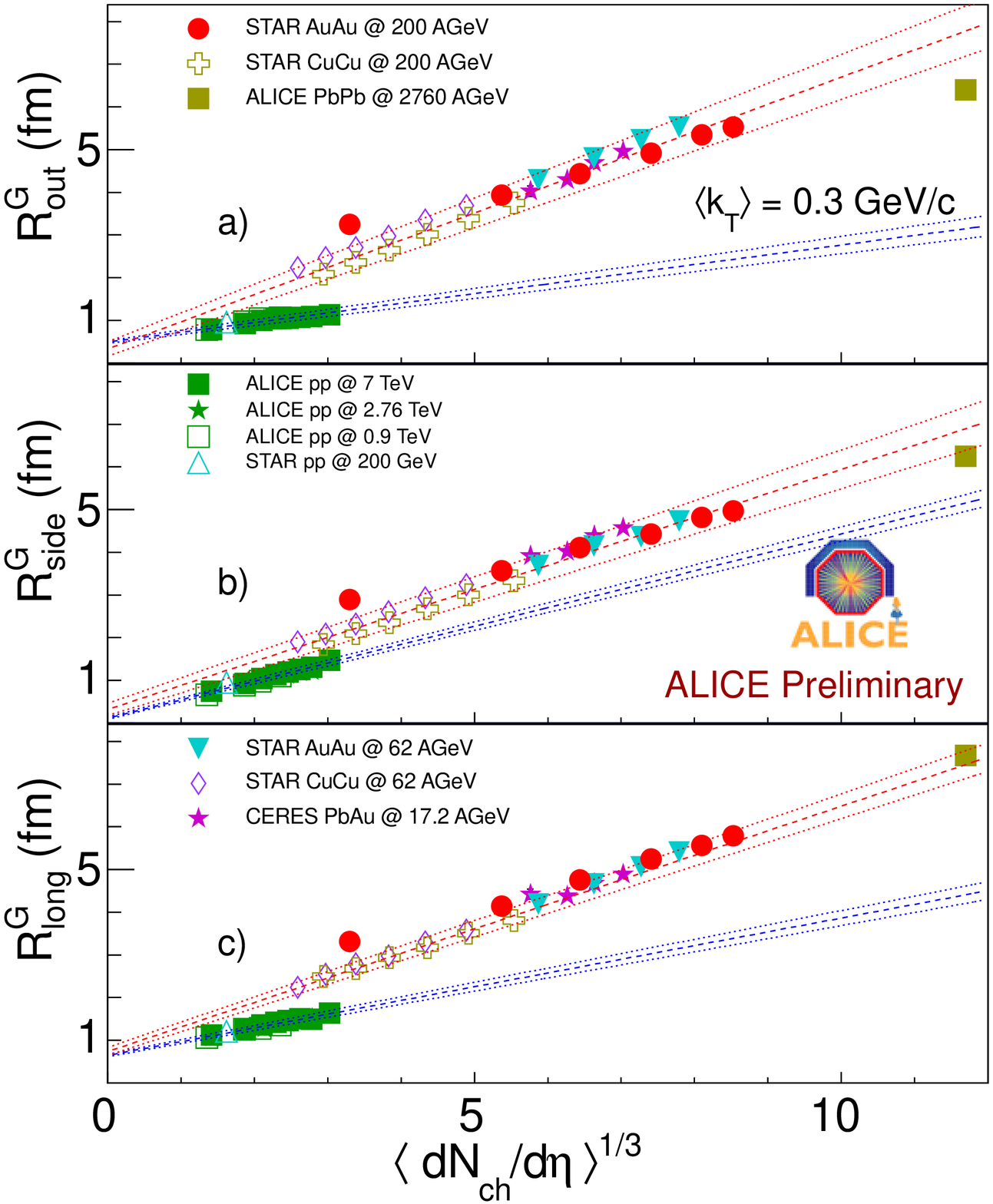}
}
\caption{Left plot: the femtoscopic radii as a function of event
  multiplicity in \pp collisions. Lines show linear fits to all data
  for two pair momentum ranges separately. Right plot: the femtoscopic
  radii as a function of event multiplicity, for \pp and the heavy-ion
  collisions at a selected $k_{T}$. Two sets of dashed lines
  represent linear fits done separately to \pp and heavy-ion data, the
  dotted lines show the uncertainty of the fit parameters.}
\label{Figfour}
\end{figure}

In Fig.~\ref{Figfour} the multiplicity dependence of the femtoscopic
radii is shown for the \pp collisions in the left panel. All collision
energies are used, for two selected $k_{T}$ ranges. The lines
represent linear fits to both $k_{T}$ sets, done separately in
each direction. For all fits $\chi^2/N_{dof}$ is below 1.0; for
$R_{out}$ at lowest $k_{T}$ the lowest multiplicity points are not
included in the fit. The monotonic increase of $R_{side}$ and
$R_{long}$ with multiplicity is observed, while for $R_{out}$ the
behavior changes from the increase at low $k_{T}$ to flattening or
decreasing at higher $k_{T}$, which can also be inferred from the radii
seen in Fig.~\ref{Figthree}. Nevertheless a linear scaling for all
femtoscopic radii with multiplicity at collision energies different by
almost an order of magnitude is seen.

In the right panel of Fig.~\ref{Figfour} the linear dependence of the
radii for \pp collisions is compared to the similar dependence from
heavy-ion collisions, including the most recent results from the ALICE
Collaboration~\cite{Aamodt:2011mr}, which extends the reach in
$\left <dN_{ch}/d\eta\right >^{1/3}$ by 50\%. The linear fit is done
separately to heavy ion data, without including the \pp points. The
quality of the fit is worse than for \pp collisions, in fact fits to central
data only gives significantly different dependence. Regardless of
which subset of data is fitted, the heavy-ion radii do not scale in
the same way than the \pp ones, in all directions. We have checked other
$k_{T}$ ranges and the same discrepancy is seen there. We conclude
that the ``universal'' scaling of femtoscopic radii with final state
multiplicity is violated by the \pp data. Such strong statement is
possible for the first time with ALICE data, because in collisions at
$\sqrt{s}=7$~TeV multiplicities comparable to the peripheral heavy-ion
collisions are produced, so direct comparison without any extrapolation
can be done. It shows that any scaling law must take into account the
initial configuration of the collision. It will be interesting to see
if a scaling law can be proposed, that would accommodate both
datasets.

The ALICE most central datapoint is significantly further in the
horizontal direction than the most central
RHIC Au-Au collision point. It can serve as a very good check of
whether the approximate linear scaling of the radii seen at lower
energies persists at the LHC. $R_{long}$ is following the linear
scaling trend from lower energies, $R_{side}$ is also in agreement,
but at the lower edge of the uncertainty in the scaling
trend. $R_{out}$ however is clearly below the trend set by lower
energies. Such a discrepancy is not foreseen in simple scaling laws, but
can be explained in hydrodynamic models~\cite{Kisiel:2008ws}. It is
therefore important to provide results for other centralities of the
Pb-Pb collisions at the LHC to determine whether the scaling is
recovered at lower multiplicities or if it is a feature of the system
evolution at higher collision energy.

\section{Kaon femtoscopy}

\begin{figure}
\centerline{
\includegraphics[width=0.6\textwidth]{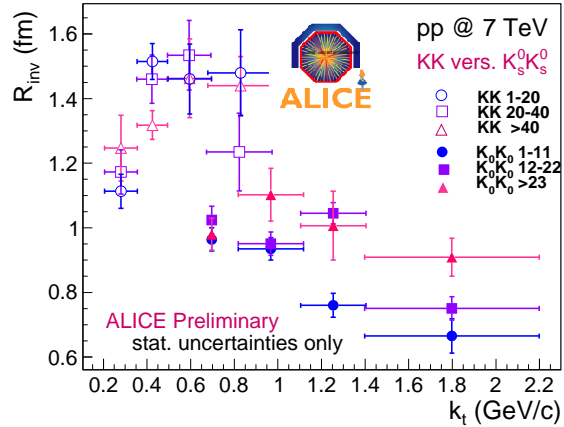}
}
\caption{Charged and neutral kaon femtoscopy invariant radius as a
  function of   event multiplicity and pair momentum. Horizontal
  error bars correspond to the $k_{T}$ bin size, vertical error bars are
  statistical only.  }
\label{Figfive}
\end{figure}

The preliminary analysis has been performed both for charged and
neutral kaons. The available statistics allowed for 1-dimensional
analysis only, multiplicity and pair momentum binning was
possible. The results are show in Fig.~\ref{Figfive}. 
 This is the first ever measurement of $\mathrm{K}^0_{s}$ correlations
in \pp collisions, and the first differential one (previous
measurements for $e^+ e^-$ and Au-Au collisions only produced a single
multiplicity and pair momentum integrated value). 
  
The study of systematic errors is still ongoing; they are presently
estimated to be 20\% for neutral kaons and 30\% for the charged ones. The
kaon analysis extends the transverse momentum range for the radii
by a factor of 3 as compared to pions. The neutral kaon
radii extend the trend seen for identical pions smoothly, while the
charged kaon radii are above the pion ones. The next step in the
analysis is to obtain the expected radii from models of \pp collisions,
especially to to compare them to predictions from models with and
without a collective stage in the evolution of the elementary
collision. 

\section{Summary}

We have presented the ALICE results of charged pion, charged kaon and
neutral kaon femtoscopy of \pp collisions at three collision
energies. The pion radii in three dimensions in LCMS scale linearly
with event multiplicity and show a decreasing trend with pair
momentum. In narrow multiplicity and pair momentum ranges they are
independent of the collision energy. The kaon results extend smoothly the
pair momentum dependence to larger pair momenta. The results were
compared to the ones from heavy-ion data, including the most central
Pb-Pb datapoint at $\sqrt{s_{NN}}=2.76$~TeV from ALICE. The \pp data
show a significantly different linear dependence than the heavy-ion
data. In addition deviations from the linear scaling are also observed
for the ALICE Pb-Pb point, in agreement with some predictions from
hydrodynamic models.


\section*{References}

\begin{thebibliography}{99}

\bibitem{Adams:2005dq}
  J.~Adams {\it et al.} [ STAR Collaboration ],
  Nucl.\ Phys.\  {\bf A757}, 102-183 (2005).
  [nucl-ex/0501009].

\bibitem{Adcox:2004mh}
  K.~Adcox {\it et al.} [ PHENIX Collaboration ],
  Nucl.\ Phys.\  {\bf A757}, 184-283 (2005).
  [nucl-ex/0410003].

\bibitem{Back:2004je}
  B.~B.~Back {\it et al.},
  Nucl.\ Phys.\  {\bf A757}, 28-101 (2005).
  [nucl-ex/0410022].

\bibitem{Arsene:2004fa}
  I.~Arsene {\it et al.} [ BRAHMS Collaboration ],
  Nucl.\ Phys.\  {\bf A757}, 1-27 (2005).
  [nucl-ex/0410020].

\bibitem{Kisiel:2008ws}
  A.~Kisiel, W.~Broniowski, M.~Chojnacki, W.~Florkowski,
  Phys.\ Rev.\  {\bf C79}, 014902 (2009).
  [arXiv:0808.3363 [nucl-th]].

\bibitem{Kopylov:1972qw}
  G.~I.~Kopylov, M.~I.~Podgoretsky,
  Sov.\ J.\ Nucl.\ Phys.\  {\bf 15}, 219-223 (1972).
  
\bibitem{Kopylov:1974uc}
  G.~I.~Kopylov, V.~L.~Lyuboshits, M.~I.~Podgoretsky,
  JINR-P2-8069
  
\bibitem{Lisa:2005dd}
  M.~A.~Lisa, S.~Pratt, R.~Soltz, U.~Wiedemann,
  Ann.\ Rev.\ Nucl.\ Part.\ Sci.\  {\bf 55}, 357-402 (2005).
  [nucl-ex/0505014].

\bibitem{Akkelin:1995gh}
  S.~V.~Akkelin, Y.~.M.~Sinyukov,
  Phys.\ Lett.\  {\bf B356}, 525-530 (1995).
  
\bibitem{Aamodt:2011mr}
  K.~Aamodt {\it et al.} [ ALICE Collaboration ],
  Phys.\ Lett.\  {\bf B696}, 328-337 (2011).
  [arXiv:1012.4035 [nucl-ex]].

\bibitem{Aamodt:2011kd}
  K.~Aamodt {\it et al.} [ ALICE Collaboration ],
  [arXiv:1101.3665 [hep-ex]], submitted to Phys. Rev. D

\bibitem{Aamodt:2008zz}
  K.~Aamodt {\it et al.} [ ALICE Collaboration ],
  JINST {\bf 3}, S08002 (2008).
 
\bibitem{Kisiel:2009iw}
  A.~Kisiel, D.~A.~Brown,
  Phys.\ Rev.\  {\bf C80}, 064911 (2009).
  [arXiv:0901.3527 [nucl-th]].

\end{thebibliography}
\end{document}